\documentclass[runningheads]{llncs}

\usepackage{amsmath}
\usepackage{graphicx}
\usepackage{subfigure}
\usepackage{array}
\usepackage{multirow}
\usepackage{todonotes}
\usepackage{amssymb}
\usepackage[utf8]{inputenc}
\usepackage[german,english]{babel}
\usepackage[english,boxed]{algorithm2e}
\usepackage{amsmath} 
\usepackage{todonotes}
\usepackage{amsfonts}
\usepackage{amssymb} 
\usepackage{graphicx}
\usepackage{placeins}
\usepackage{algorithmic}
\usepackage{subfigure}
\usepackage[titletoc]{appendix}
\usepackage{color}
\usepackage{chapterbib}
\usepackage{tcolorbox}
\usepackage{pdfpages}
\usepackage{hyperref}
\usepackage{array}
\usepackage{mathdots}
\usepackage{setspace}
\usepackage{authblk}
\usepackage{pdfpages}

\DeclareMathOperator*{\argmin}{arg\,min}

\newcommand{\field}[1]{\ensuremath{\mathbb{#1}}}

\newcommand{\R}{\field{R}}

\newcommand{\norm}{\lVert}
\begin{document}

\title{Blood vessel segmentation
in en-face OCTA images: a frequency based method}
%
%
\author{Anna Breger\inst{1}\thanks{Corresponding author, E-mail: anna.breger@univie.ac.at} \and 
Felix Goldbach \inst{2} \and
Bianca S. Gerendas \inst{2} \and Ursula Schmidt-Erfurth \inst{2} \and
Martin Ehler \inst{1} }
%

\authorrunning{Breger \textit{et al.}}

\institute{Department of Mathematics, University of Vienna, Austria \and Department of Ophthalmology and Optometry, Medical University of Vienna, Austria}




\maketitle              

\begin{abstract}
Optical coherence tomography angiography (OCTA) is a novel noninvasive imaging modality for visualization of retinal blood flow in the human retina. Using specific OCTA imaging biomarkers for the identification of pathologies, automated image segmentations of the blood vessels can improve subsequent analysis and diagnosis. We present a novel segmentation method for vessel density identification based on frequency representations of the image, in particular, using so-called Gabor filter banks. The algorithm is evaluated qualitatively and quantitatively on an OCTA image in-house data set from $10$ eyes acquired by a Cirrus HD-OCT device. Qualitatively, the segmentation outcomes received very good visual evaluation feedback by experts. Quantitatively, we compared the resulting vessel density values with automated in-built values provided by the device. The results underline the visual evaluation. For the evaluation of the FAZ identification substep, manual annotations of $2$ expert graders were used, showing that our results coincide well in visual and quantitative manners. Lastly, we suggest the computation of adaptive local vessel density maps that allow straightforward analysis of retinal blood flow in a local manner.  
\end{abstract}

\section{Introduction}
\label{sec:introduction}

Optical coherence tomography angiography (OCTA) is a novel technique for noninvasive, rapid imaging and quantification of retinal blood flow and has expanded the diagnostic toolbox in ophthalmology. OCTA enables the visualization of the retinal and choroidal vascular networks (by creating a 3D angiogram) based on detection of red blood cell movement.  These vascular networks and their specific biomarkers can be investigated individually by creating a $2$D top view image (called en-face image) of the chosen slab. OCTA en-face images are mostly 8-bit grayscale images with very small grayish-dark vessels and large whitish-bright retinal vessels. Dark pixels correspond to no retinal blood flow. For subsequent automated analysis of some biomarkers it is important to transform these grayscale en-face images into segmented binary images, where black pixels correspond to the absence of blood flow and white pixels indicate blood flow.

Several studies have demonstrated the importance of OCTA imaging biomarkers, e.g. vessel density or the foveal avascular zone (FAZ), for early detection and
monitoring of several retinal diseases but also systemic diseases like diabetes \cite{10.1167/iovs.18-24142}, acute renal injury \cite{renal}, cardiovascular disease \cite{10.1371/journal.pone.0194694,10.1167/iovs.18-24090} and arterial hypertension \cite{hypertension}. A correct segmentation of the retinal capillary networks is critical for the computation of such  markers and its subsequent clinical applicability. Unfortunately, it is not feasible to manually annotate the vessel architecture in a bigger amount of OCTA en-face images and therefore a tool for automated segmentation is much-needed. Moreover, consistent segmentations of OCTA data sets can subsequently be evaluated more easily by humans and machines, improving diagnosis and therapy of patients. One main challenge therein is to obtain a representation of the vessel architecture that still maintains the actual vascular network.

Finding good, and especially generalizable (across devices), automated segmentation tools for retinal vessels in OCTA scans remains an open problem in clinical routine and research \cite{segocta, 9284503}. Some OCTA manufacturers provide their own binarization algorithms for an expanded OCTA analysis but these are device and platform dependent and have their own limitations in particular regarding vessel density. Deep learning is nowadays widely used for diverse medical imaging problems. The huge drawback is the need for a big amount of annotated data in order to yield accurate results, and especially, to avoid overfitting and allowing generalization to new data. E.g. in \cite{10.1167/tvst.9.13.5} an automated vessel segmentation algorithm has been proposed based on deep learning with annotated OCTA data. They use patches derived from $11$ OCTA scans to train convolutional neural networks. 

We will present an algorithm that is independent from annotated data, segmenting OCTA en-face images based on frequency representations. In particular, we use so-called Gabor filters that are known to be closely connected to the perception of mammalian brains \cite{10.1109/34.41384}, and have been successfully applied in texture analysis and denoising \cite{531803, gabordenoising}. Moreover, they have been successfully used in medical applications \cite{gabormed}, and in particular in segmentation tasks of OCT images \cite{gaboroct, Breger:2017aa, octvesselgabor}. Beyond that, a novel Gabor OCTA imaging algorithm \cite{gaborocta} was proposed, suggesting to include Gabor filters in the actual OCTA processing to receive the en-face vascular images. 

 We use a comprehensive Gabor filter bank that allows enhancement of selected frequency ranges in the new image representations, further described in Section 2. These preprocessed images enable successful local and global thresholding for the identification of small and big vessels. The last step of the algorithm is the segmentation of the FAZ and is approached by variational energy minimization, the so-called Potts problem \cite{Storath:2014ud}, yielding a piecewise constant representation of the image that enables direct identification of the area. In Section 3 the algorithm is evaluated quantitatively and qualitatively on OCTA in-house data from $10$ eyes with various retinal diseases. Our vessel segmentations received positive qualitative feedback from expert clinicians. In the quantitative evaluation, we compare the resulting vessel density (VD) with corresponding in-built values returned by the OCTA device. Our segmentations are very reliable and accurate with a mean absolute discrepancy of $1.58\pm 1.08$\% compared to the device. For the evaluation of the FAZ identification we compare the segmentations with $2$ annotations of expert graders serving as our ground truth (GT). The mean absolute differences for the descriptors \textit{area}, \textit{perimeter} and \textit{circularity} are $0.0068 \pm 0.0043 mm^2$, $0.1632 \pm 0.1409 mm$, $0.1242 \pm 0.1136$,  respectively. Moreover, we provide Bland-Altman plots for the statistical analyses of the $3$ quantities as well as the dice similarity coefficient \cite{10.2307/1932409} of our segmentations and the GT. Lastly, we introduce local vessel density maps, which allow a direct interpretation of the blood flow in regions of any desired size. 

\section{Methods}
\label{sec:methods}
The first important step is the preprocessing of the data with filters derived from a so-called Gabor filter bank.

\begin{figure}
\centering
    \subfigure[$I_1$: High frequencies]{\includegraphics[width=0.31\textwidth]{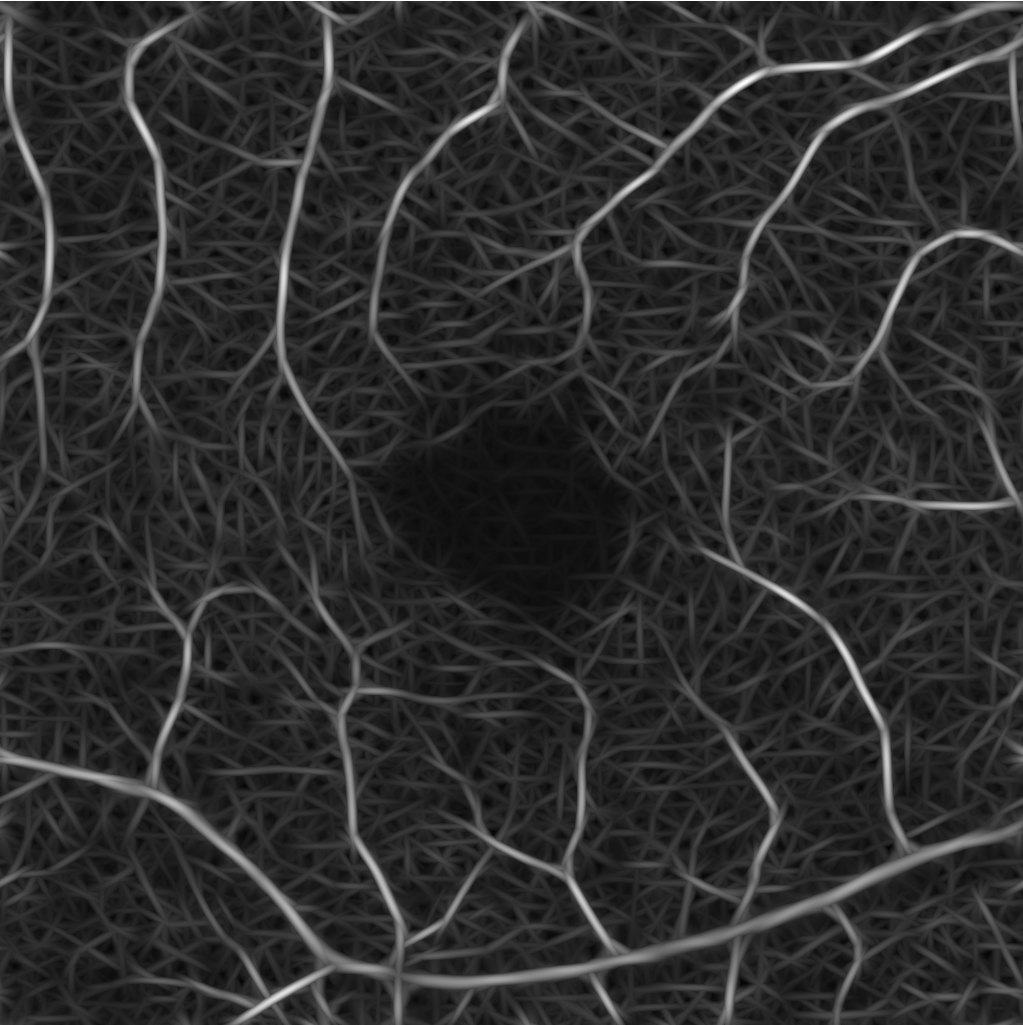} }
    \subfigure[$I_2$: Medium frequencies]{\includegraphics[width=0.31\textwidth]{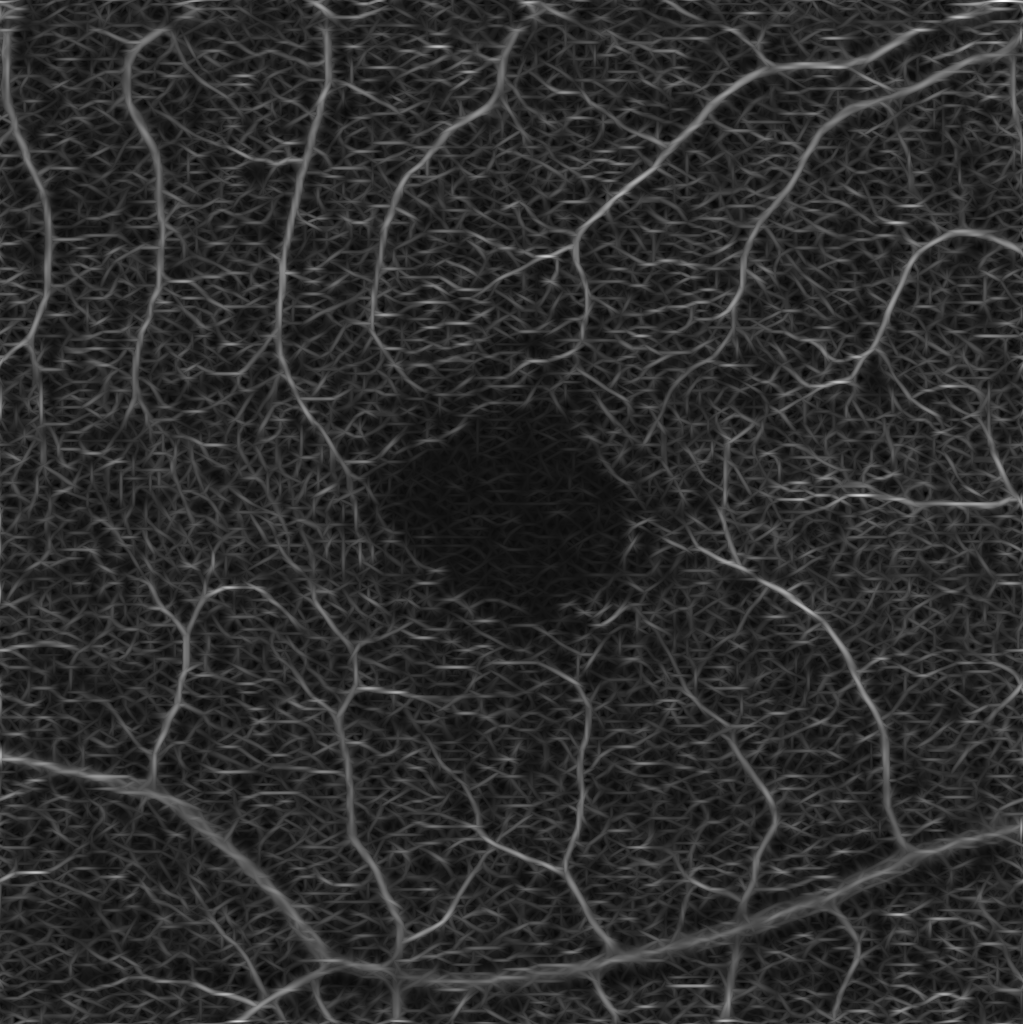} } 
    \subfigure[$I_3$: Low frequencies]{\includegraphics[width=0.31\textwidth]{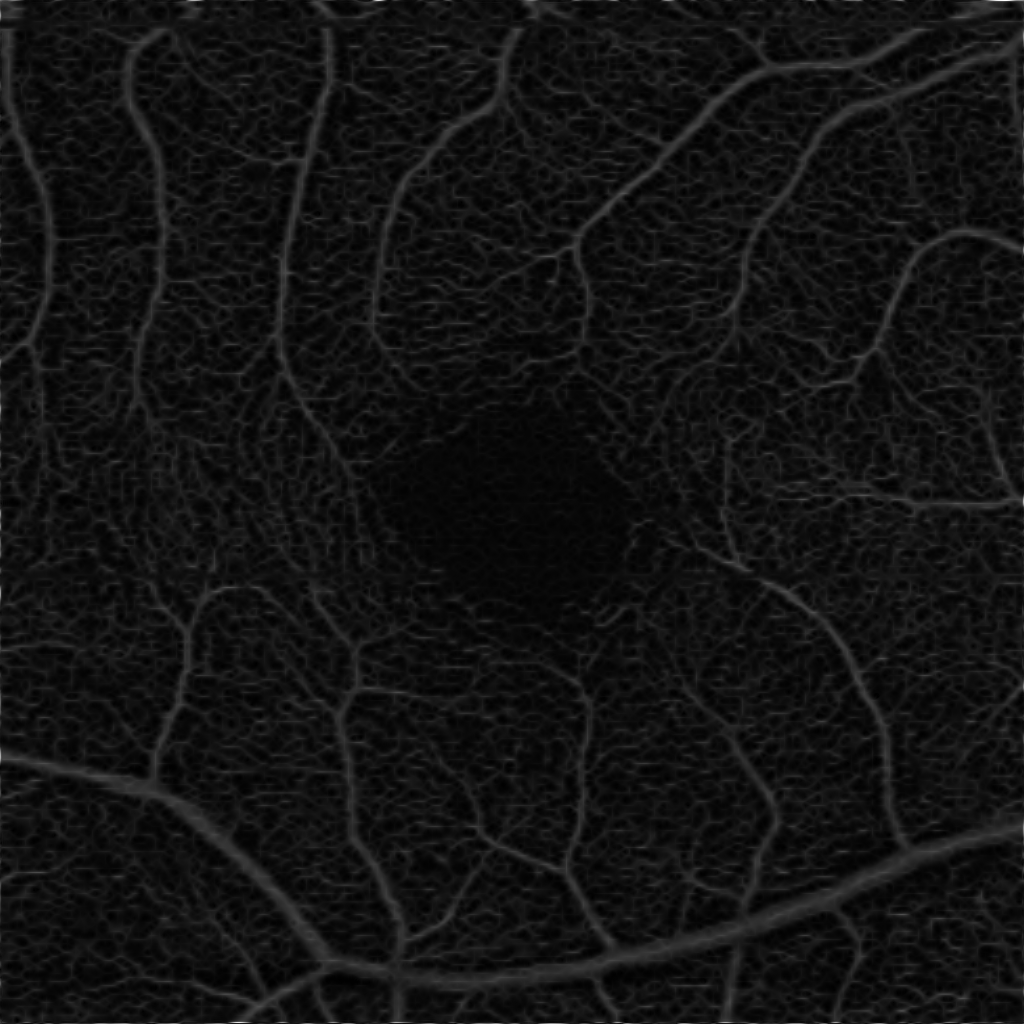} }
    \caption{New image representations are obtained by the convolution of the original image $I$ with Gabor filters. The images $I_1, I_2, I_3$ correspond to the $3$ chosen frequency ranges defined by $\{\omega_i \}_{i=1}^3$.}
    \label{gaborfrequencies}
\end{figure}

A continuous Gabor filter can be written as 
\begin{equation}\label{eq}
    g_{\theta,\omega}(x_1,x_2) := \frac{\omega^2}{\pi\sigma_1 \sigma_2} \  e^{-\omega^2\big(\frac{(x_1 \cos\theta + x_2 \sin\theta)^2}{\sigma_1^2}+\frac{(-{x_1} \sin\theta + x_2 \cos\theta)^2}{\sigma_2^2}\big)}e^{2\pi i \omega ({x_1} \cos\theta + {x_2} \sin\theta)},
\end{equation}
for some chosen frequency $\omega$, orientation $\theta$ and $\sigma_1$, $\sigma_2$ correspond to the spatial widths of the filter, which is then applied to an image. Here, we choose the parameters as suggested in \cite{531803} and use $3$ frequency scalings $\{\omega_i \}_{i=1}^3$ and $6$ orientations $\{\theta_j \}_{j=1}^6$ to create our Gabor filter bank. The $3 \cdot 6 = 18$ computed Gabor filters are then convolved with the original image $I \in \R^{m \times n}$ to derive the new image representations in $\R^{m \times n}$. Next, for fixed $\omega_i$, we take the pixel-wise maximum in the images corresponding to the $6$ orientations $\{\theta_j \}_{j=1}^6$, leading to $3$ images $I_1, I_2, I_3$ with different emphasized frequency ranges, see Figure \ref{gaborfrequencies}. These new representations are our basis for the further processing.

\begin{figure}
    \subfigure[Original image] {\includegraphics[width=0.49\textwidth]{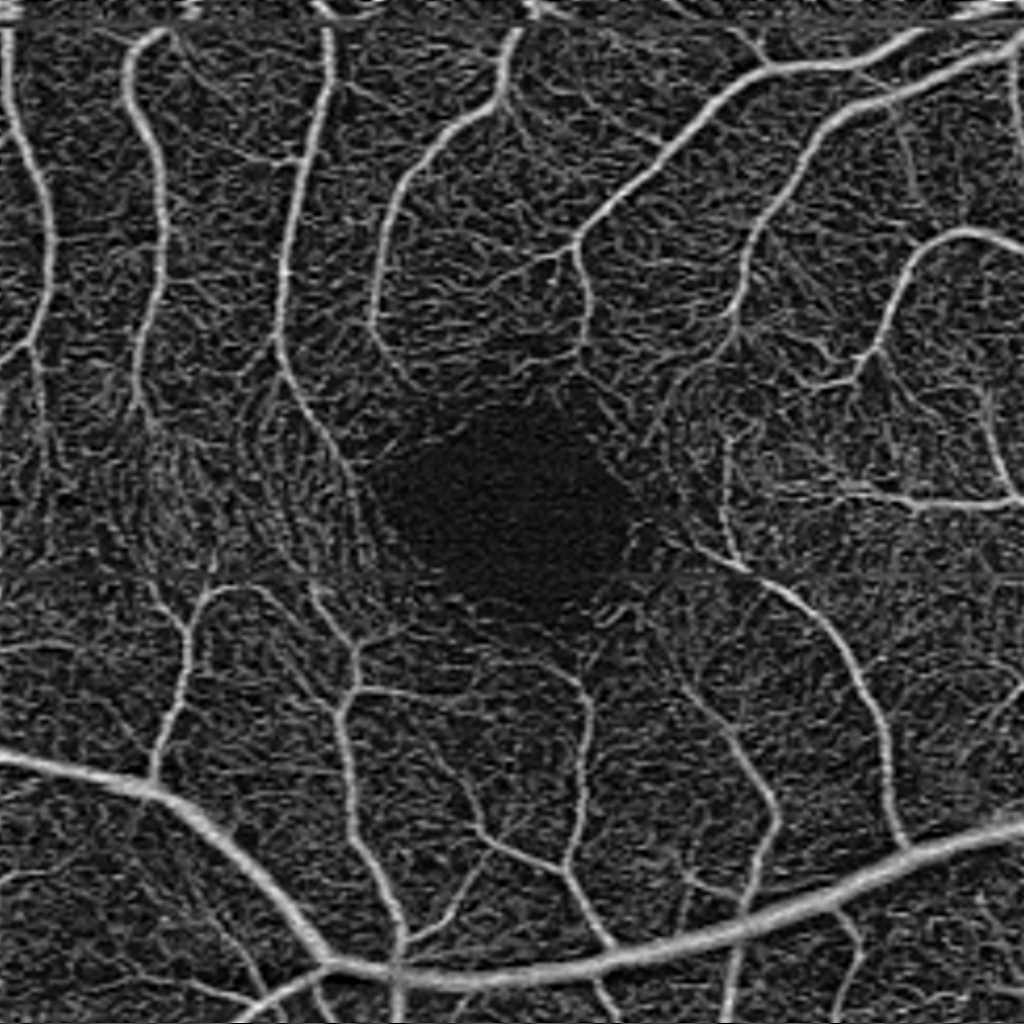} }
    \subfigure[Final segmentation ]{\includegraphics[width=0.49\textwidth]{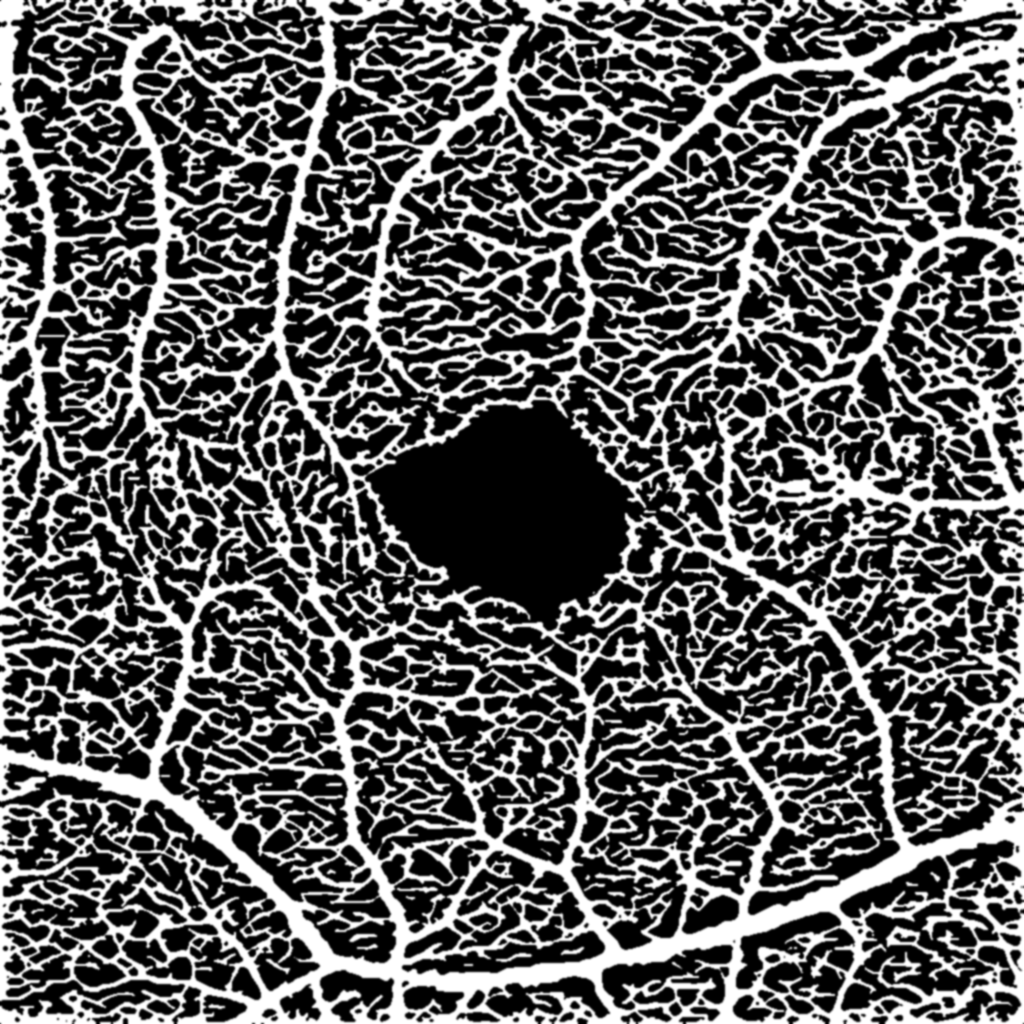} }
    \caption{The combined substeps described in Sections \ref{vessels} and \ref{fazid} yield the final image.}
    \label{final}
\end{figure}

\noindent
Our image segmentation is divided into three subtasks: the identification of small and big vessel and the FAZ. The final result, see Figure \ref{final}(b), is obtained by uniting the binary images with the small and big vessels and subsequently using the FAZ segmentation as a mask that excludes that area. 

\subsection{Identification of small and big vessels}\label{vessels}
For the image processing steps that focus on the identification of the small and big vessels, we will use the image representation $I_2$ that emphasizes the medium frequencies. These frequencies correspond best to the vessels we aim to identify and therefore the new representation reveals a more clear structure of the relevant characteristics. We smooth the image $I_2$ by convolution with a Gaussian filter 
\begin{equation}\label{gauss}
    h(x_1,x_2) := \frac{1}{2 \pi \sigma^2} \exp^{\big(-\tfrac{x_1^2+x_2^2}{2 \sigma^2}\big)},
\end{equation}
with standard deviation $\sigma = 1$, in the following referring to this smoothed version as $I_2$.


In order to identify the smaller vessels, we compute a threshold from local and global statistics of the preprocessed image. In particular, we apply two order-statistic filters to the image: the local one takes the maximum of a $3 \times 3$ neighborhood, the global one the median of a $30 \times 30$ neighborhood. After the filters are applied to $I_2$, we obtain two new images. Finally, we threshold the image $I_2$ pixel-wise by the pixel-wise mean of the two images, yielding a binary segmentation of the small vessels.  

To identify the big vessels, we apply the well known thresholding method by Otsu (cf. \cite{4310076}) to $I_2$.

\subsection{Identification of the FAZ}\label{fazid}
For the identification of the FAZ, we will use the image that is obtained by taking the pixel-wise maximum over all frequency ranges $\{\omega_i \}_{i=1}^3$ and orientations $\{\theta_k \}_{j=1}^6$, i.e. the pixel-wise maximum of the images $I_1, I_2$ and $I_3$, followed by a Gaussian filter \eqref{gauss} with $\sigma = 2$. In this task particular frequencies do not play a substantial role since there are no vessels to be identified in the FAZ. 
\begin{figure}
    \subfigure[Smoothed Gabor representation]{\includegraphics[width=0.49\textwidth]{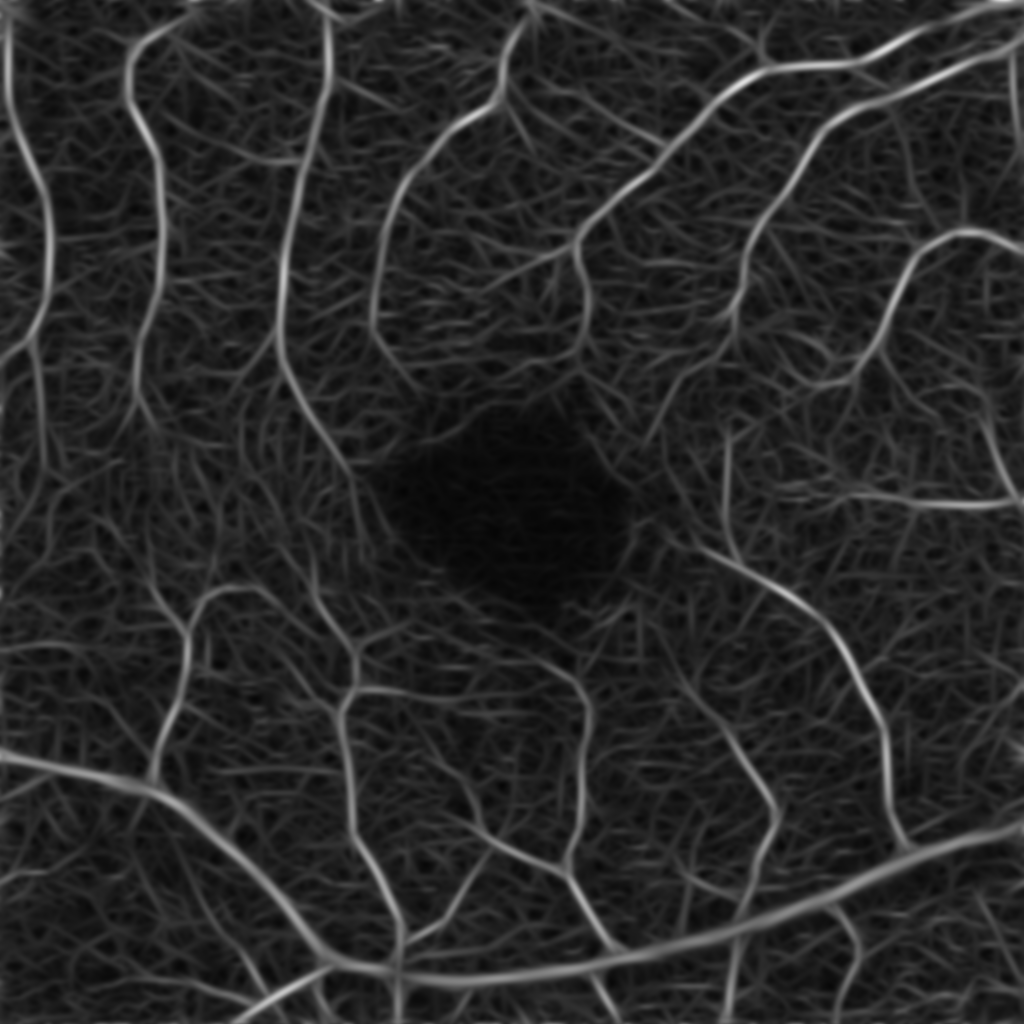} }
    \subfigure[Piecewise constant image ]{\includegraphics[width=0.49\textwidth]{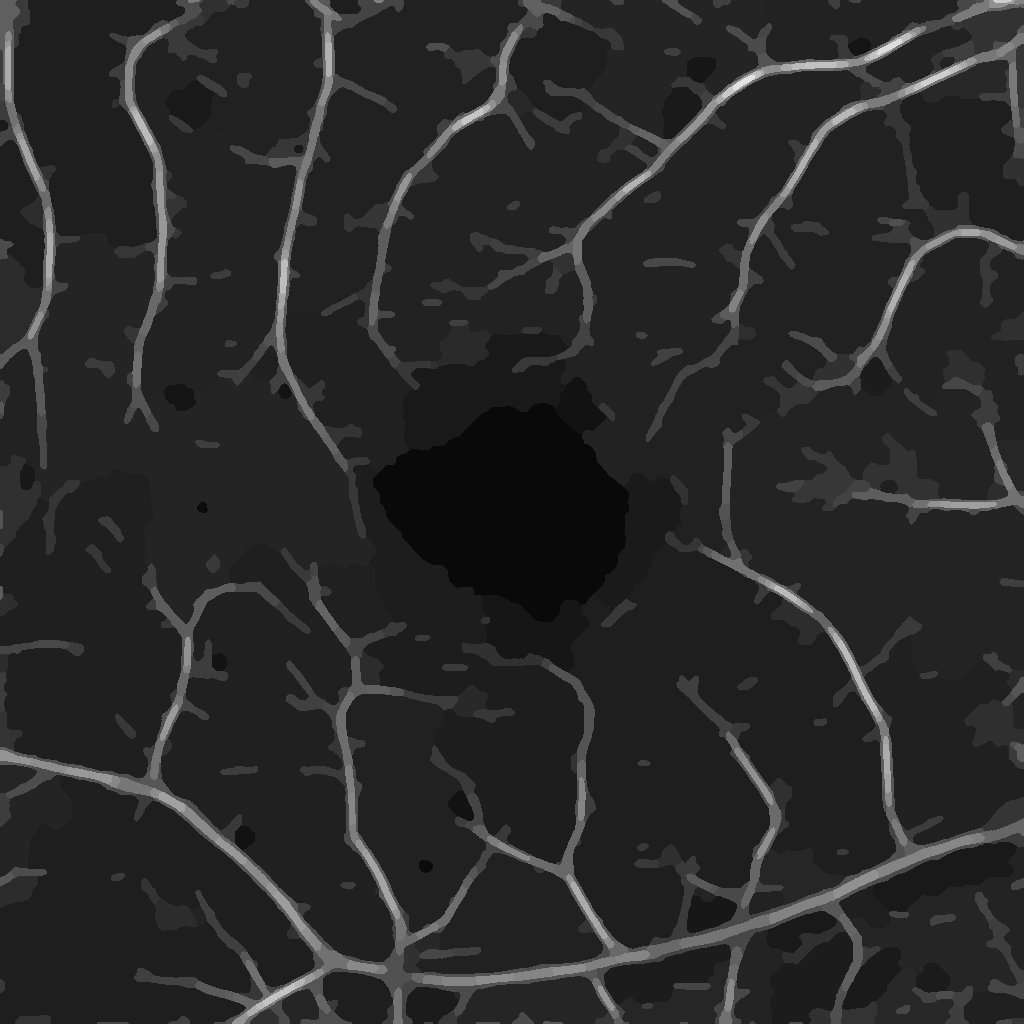} } 
   \caption{Image representations for the FAZ identification}
    \label{gaborall}
\end{figure}
To segment the FAZ region, we use a variational energy minimization problem to obtain a piecewise constant representation of the image. We employ the so-called Potts model, which can be formulated as follows: For a given  initial image $z:\Omega \rightarrow \R^r$ on a continuous region $\Omega \subset \R^2$ and $\gamma>0$, find the piecewise constant minimizer of

\begin{equation}\label{potts} 
\argmin_u \gamma \cdot \norm \nabla u \norm_0 + \int_\Omega \big(u(x) - z(x)\big)^2 dx,
\end{equation}
where $u : \Omega \rightarrow \R^r$ denotes a bounded piecewise constant function and $\norm \nabla u \norm_0$ the total boundary length of its partitioning, cf. \cite{Mumford:1989uq,10.1109/34.969114,Storath:2014ud} for details. The second term provides approximation to the original data and the first term enforces the partitioning, where the parameter $\gamma$ controls the balance between the two penalties. We use the fast strategy and code presented in \cite{Storath:2014ud} for solving a discretization of \eqref{potts} to receive the piecewise constant image representation, see Figure \ref{gaborall}(b).
The FAZ is then identified by choosing the darkest segment in the central part of the image.  

\section{Data, Results and Discussion}
\label{sec:data}
OCTA examinations were performed with a Cirrus HD-OCT (Angioplex) by an experienced ophthalmologist according to a standardized protocol as part of an OCTA device comparison study in 2020 at the Department of Ophthalmology and Optometry at the Medical University of Vienna. The study was approved by its ethics committee (EK\#1437/2019) and was conducted in accordance with the Declaration of Helsinki. $3\times3$ mm OCTA scans with various retinal diseases were recorded and en-face images of the superficial vascular network were exported for this investigation. All enrolled patients provided written informed consent for analysis and publication of their OCTA images. For the evaluation we chose $10$ eyes with varying diseases and similar image quality, where $5$ patients were used to visually fine-tune the parameters of the algorithm. 

The described data can be returned by the device with corresponding VD values in percentage. Our decision to evaluate the full segmentations concerning their corresponding VD values is twofold. First, VD is of main interest in the segmentation and has been identified as a very important biomarker for vascular diseases in OCTA \cite{rep,10.1167/iovs.15-18904,vdi}. Secondly, it is highly time-consuming and extremely challenging to annotate small blood vessels in OCTA images in a consistent and feasible manner that would allow an accurate pixel-wise comparison. The device itself does not return binary blood vessel segmentations. Therefore, the VD values provide us the baseline for quantitative evaluation. The substep of FAZ identification is evaluated by comparison to manual annotations of $2$ expert graders. Additionally, the device returns a FAZ segmentation, which is also included in the evaluation. 


\subsection{Qualitative results}\label{qual}
\begin{figure}
    \subfigure{\includegraphics[width=0.24\textwidth]{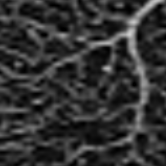}}
    \subfigure{\includegraphics[width=0.24\textwidth]{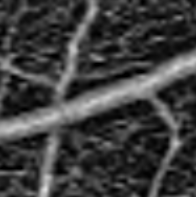}}
    \subfigure{\includegraphics[width=0.24\textwidth]{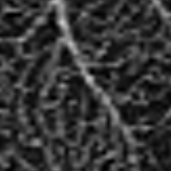}} 
    \subfigure{\includegraphics[width=0.24\textwidth]{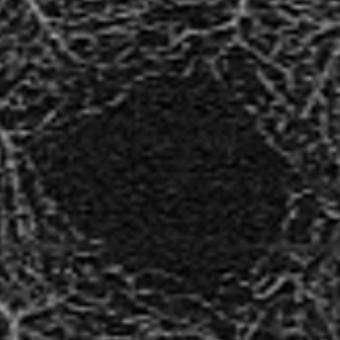}} 
    \\
    \subfigure{\includegraphics[width=0.24\textwidth]{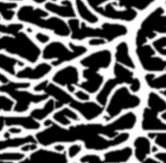}}
    \subfigure{\includegraphics[width=0.24\textwidth]{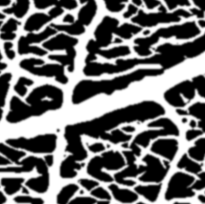}}
    \subfigure{\includegraphics[width=0.24\textwidth]{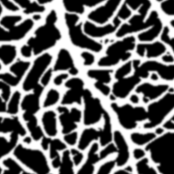}} 
    \subfigure{\includegraphics[width=0.24\textwidth]{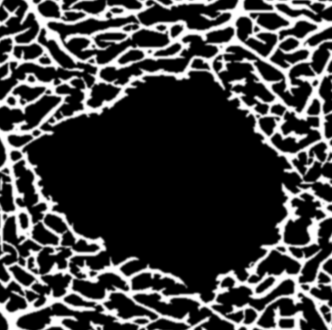}} 
    \caption{Originals (top) and segmentations (bottom) from the proposed method}
    \label{details}
\end{figure}
\subsubsection{Vessel Density}
The visual results of the segmentation are very good: small vessels are identified accurately, branches are preserved and the ratio of the vessel diameter is pictured realistically. Furthermore, small non-perfusion areas around larger vessel and vascular angles are also depicted truthfully. See Figure \ref{details} for a visualisation of some details. 

\subsubsection{FAZ identification}
\begin{figure}
\centering
    \subfigure[FAZ - annotation]
    {\includegraphics[width=0.3\textwidth]{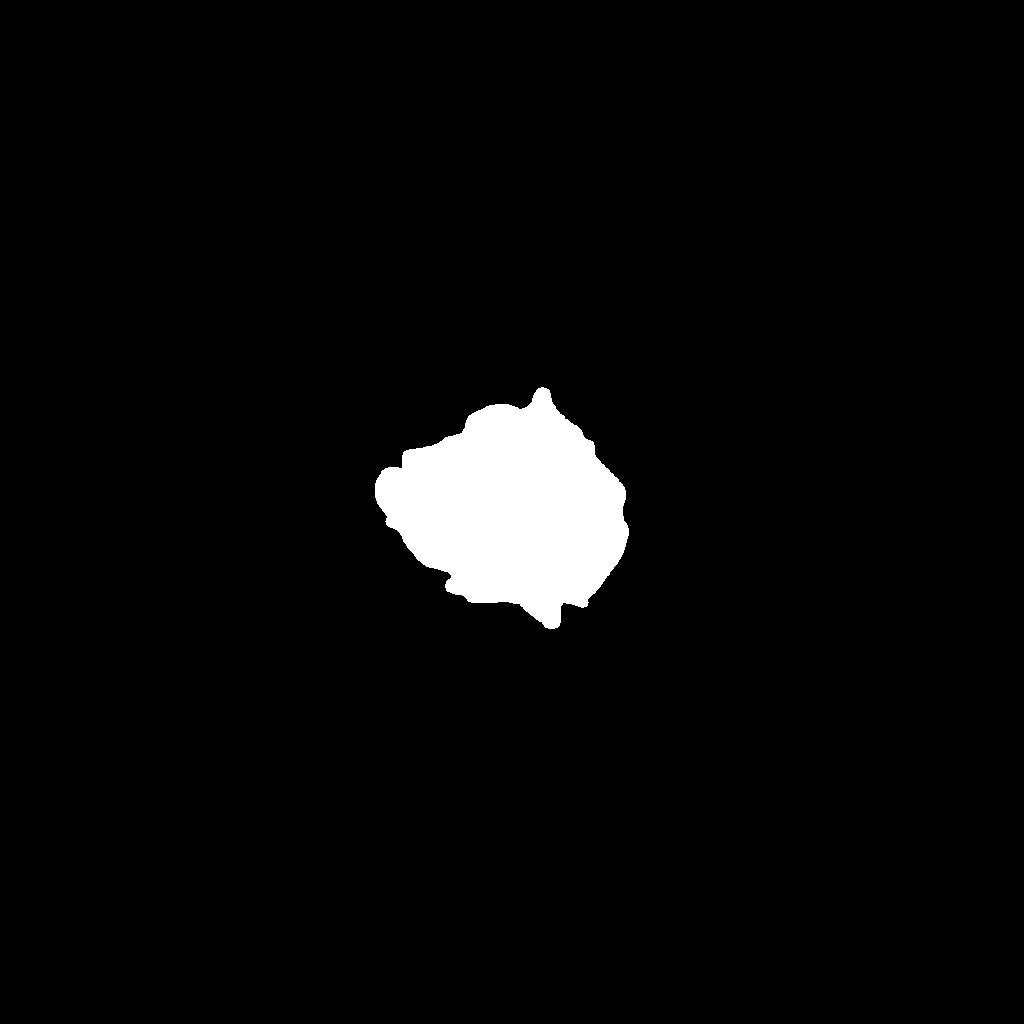}} \quad
    \subfigure[FAZ - device]
    {\includegraphics[width=0.3\textwidth]{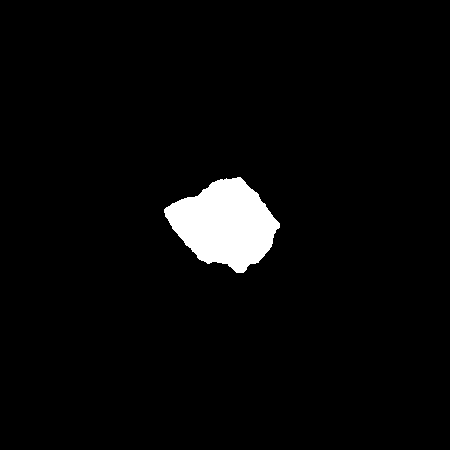}} \quad
    \subfigure[FAZ - algorithm] {\includegraphics[width=0.3\textwidth]{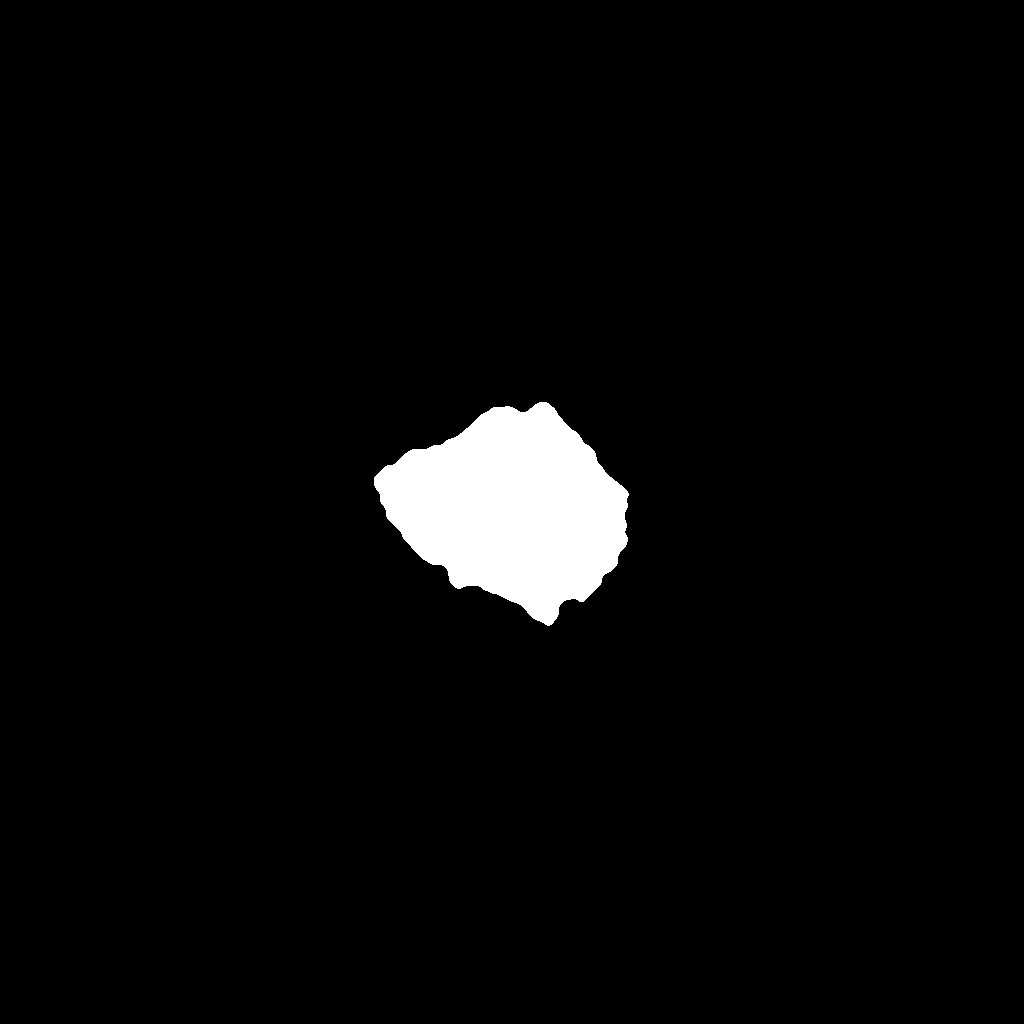}}
    \caption{FAZ identified by an expert grader(a) from the device(b) and our algorithm(c).} 
    \label{faz}
\end{figure}

In Figure \ref{faz} we can see the FAZ annotated manually by one expert grader (a), returned from the device (b) and the result from our algorithm (c). We observe that the manual annotation and the automated one from the device differ in their main characteristics and that our segmentation contains characteristics of both, e.g. shape and details of the boundary. 

\subsection{Quantitative results}
\subsubsection{Vessel Density (VD)} 
\begin{figure} 
\centering
    \subfigure[ETDRS grid (device)] {\includegraphics[width=0.46\textwidth]{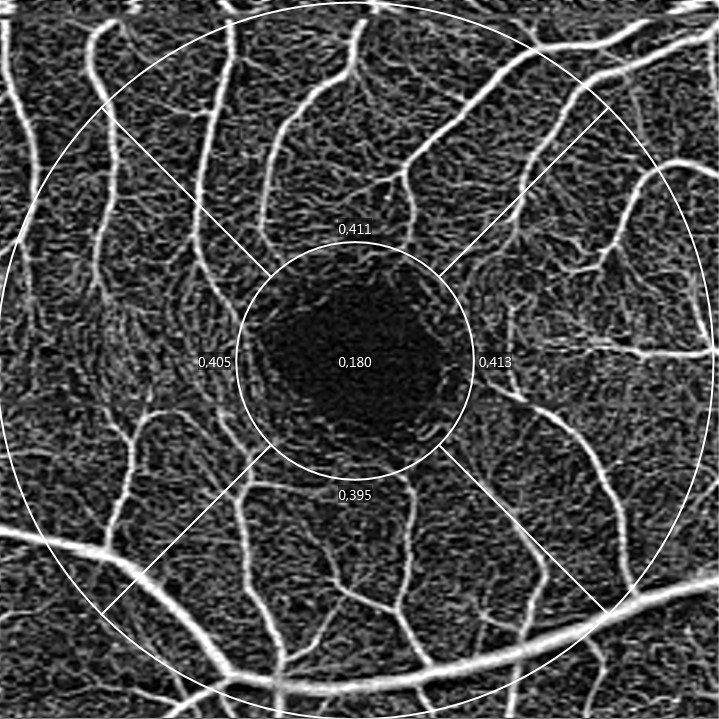}} \quad
    \subfigure[VD values (device)] {\includegraphics[width=0.46\textwidth]{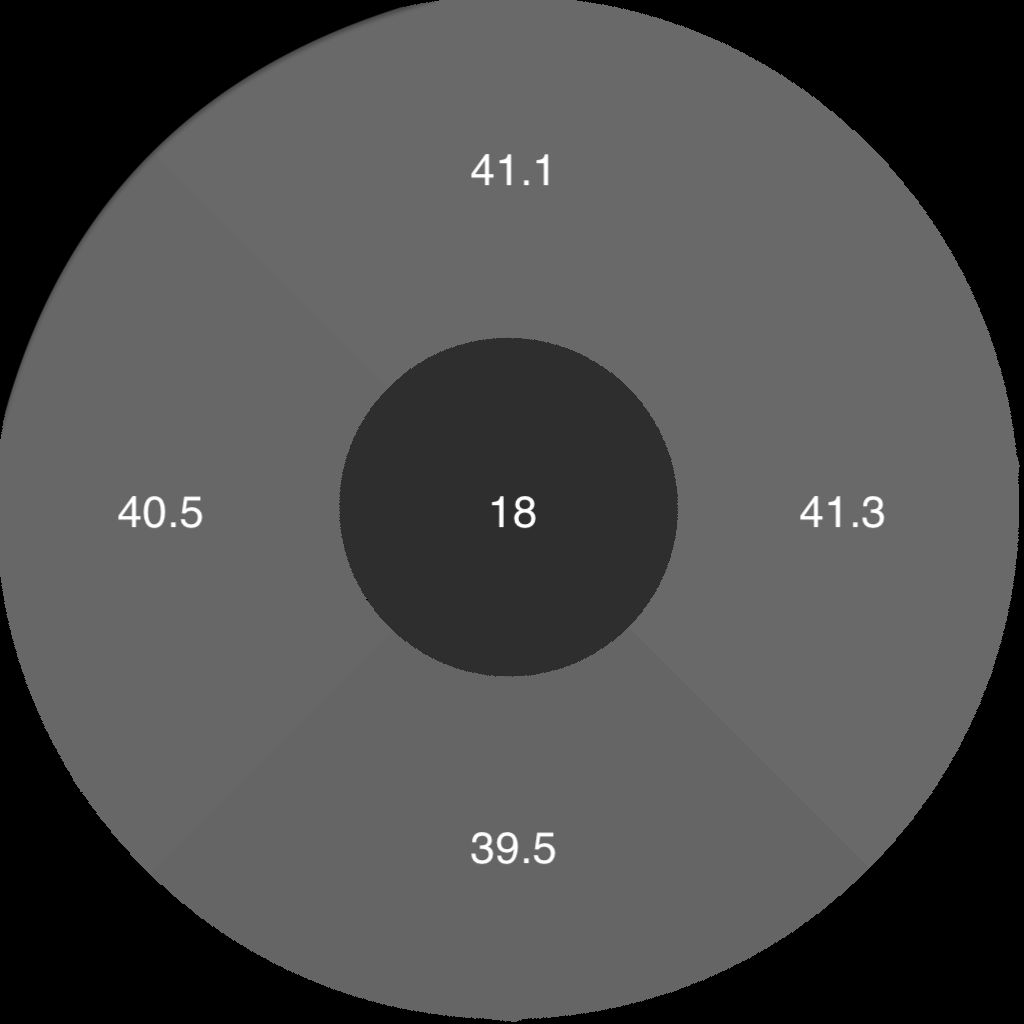} }
    \caption{VD is computed by the device in five areas of the $3$mm: $1$mm and temporal, superior, nasal and inferior quadrant. The left image shows the corresponding ETDRS grid and the right image shows the VD values provided by the device.}
    \label{grid}
\end{figure}

The OCTA device computes VD values with an in-built algorithm and returns it in $\%$ for defined areas, i.e. the number of pixels containing blood flow (all pixels $> 0$) divided by the number of all pixels in that area. The typical areas of interest are defined by the inner areas of an ETDRS grid, see Figure \ref{grid}(a). 
The grid for the 3x3mm scans divides the retina into five large regions bound by two rings: a central foveal circle with 1mm diameter and an inner parafoveal ring around it with $3$mm diameter divided into four quadrants; temporal, superior, nasal and inferior. 
For these areas the VD values are provided by the OCTA device. We observed that the vessel identification of the device is quite inaccurate concerning particular vessels, i.e. more local areas, see Figure \ref{grid} for an example: the vessel density in the inferior quadrant is the lowest ($39.5\%$) although the biggest vessel appears there and simple numerical experiments as well as visual evaluation indicate that this is not accurate. Due to unrealiable values from the in-built device software for big vessels, we decided to compare the VD values more globally within the $3$mm parafoveal ring, in the following referred to as the $3$-$1$ ring.

\begin{figure}
\centering
    \subfigure[Bland-Altman Plot]{\includegraphics[width=0.5\textwidth]{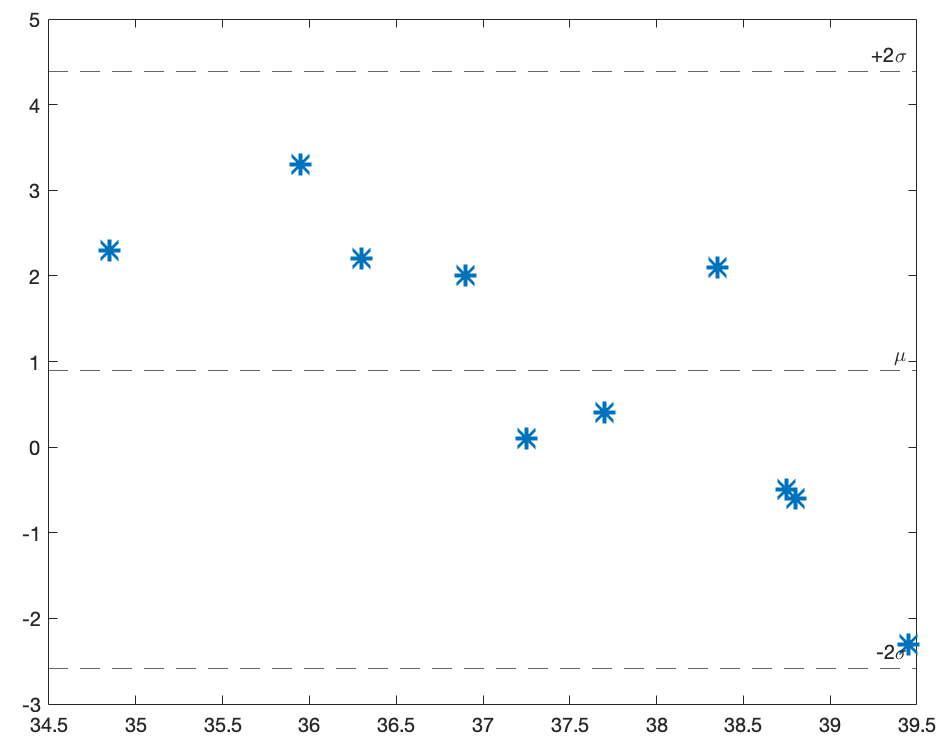}} \qquad
        \subfigure[Skeleton map] {\includegraphics[width=0.4\textwidth]{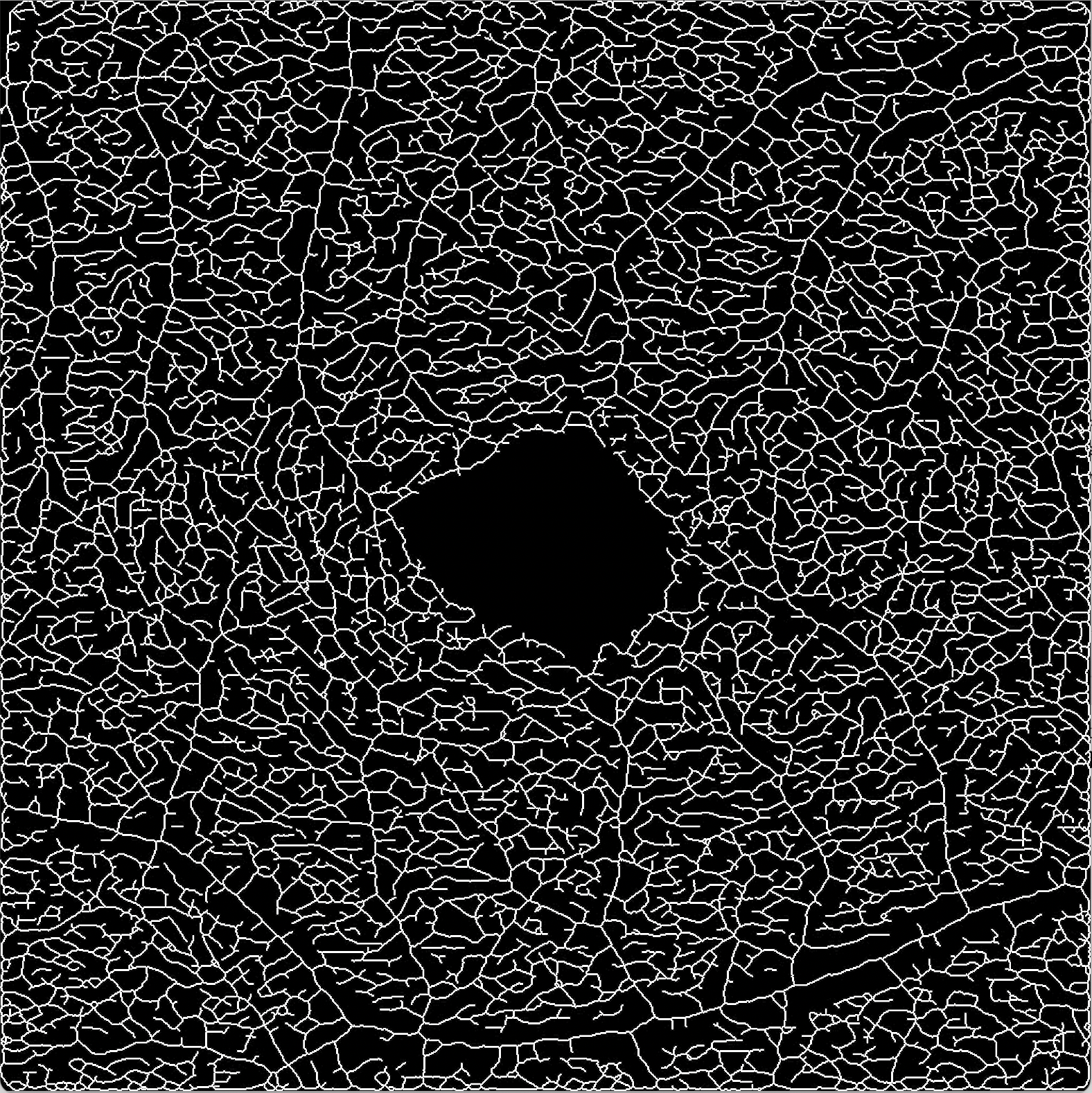}}
        \caption{Bland-Altman Plot (left) regarding VD values (in $\%$) for $10$ patients provided by our algorithm versus the values returned by the OCTA device itself.  The $x$-axes denotes the mean of the observed values by the two methods, the $y$-axes the differences. The right image (b) corresponds to a vessel skeleton map derived from our segmentation.}
    \label{VDbland}
\end{figure}

The Bland–Altman plot, see Figure \ref{VDbland}, shows a good agreement between the results of our algorithm and the device measurements for the VD values of all $10$ eyes. Furthermore, the mean absolute differences between our algorithm and the device VD were $1.58 \pm 1.08\%$ and the largest discrepancy is $3.3\%$. These results are very promising since VD identification is highly complicated, e.g. the best model in \cite{10.1167/tvst.9.13.5} leads to a mean VD difference of $6\%$ in their data with manual annotations. 

\subsubsection{FAZ identification}
To evaluate our FAZ segmentations quantitatively we first compare the common descriptors \textit{area}, \textit{perimeter} and \textit{circularity} \cite{articlefaz}. These values are also provided by the OCTA device itself and furthermore, we obtained independent manual FAZ annotations from three expert graders. Visual and statistical analyses showed that the values of the two graders and the OCTA device correlate highly, whereas the annotations of the third grader yield very different values. E.g., the Pearson correlation coefficients regarding \textit{perimeter} between the third grader and the two remaining graders, as well as the device's output, are respectively $0.44, 0.48$ and $0.46$. All other combinations yield at least $0.88$. Therefore, we decided to discard the values of that grader and use the mean values of the $2$ remaining expert graders as ground truth (GT). 

\begin{figure}
\centering
\subfigure[FAZ segmentation (algorithm): \textit{area }$(mm^2$)]{\includegraphics[width=0.45\textwidth]{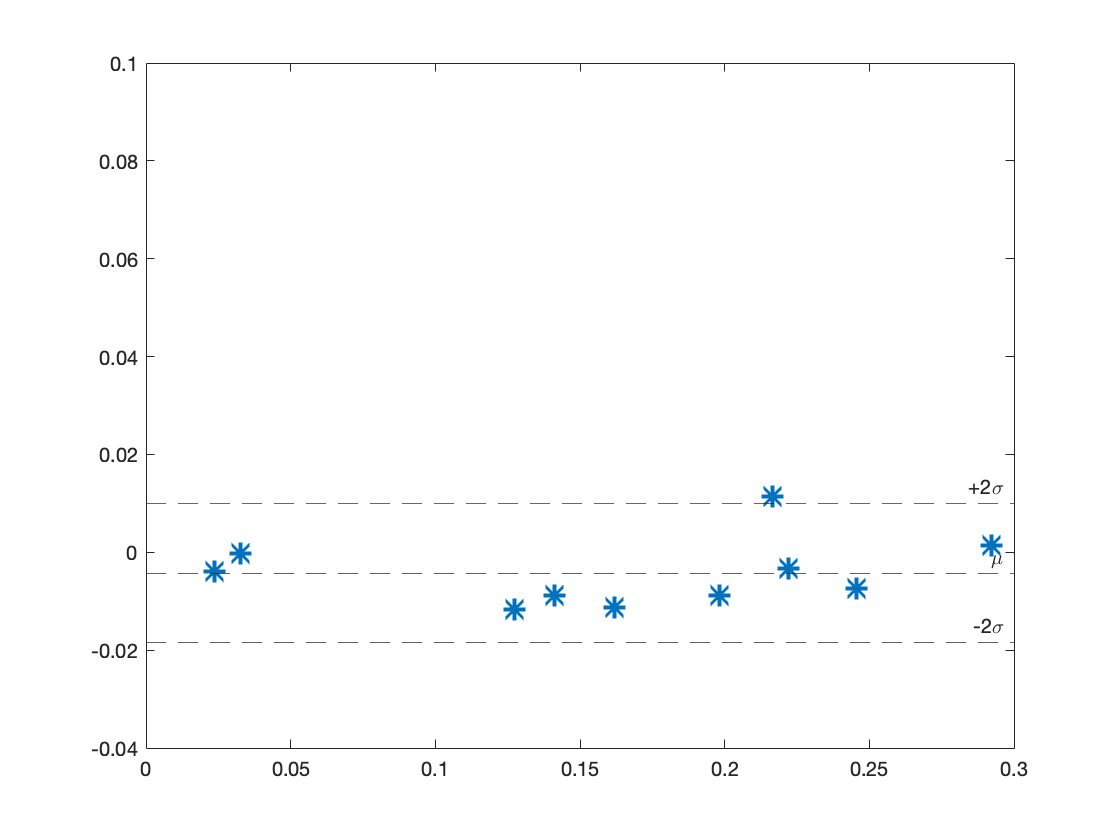}} 
\subfigure[FAZ segmentation (device): \textit{area }$(mm^2)$]{\includegraphics[width=0.45\textwidth]{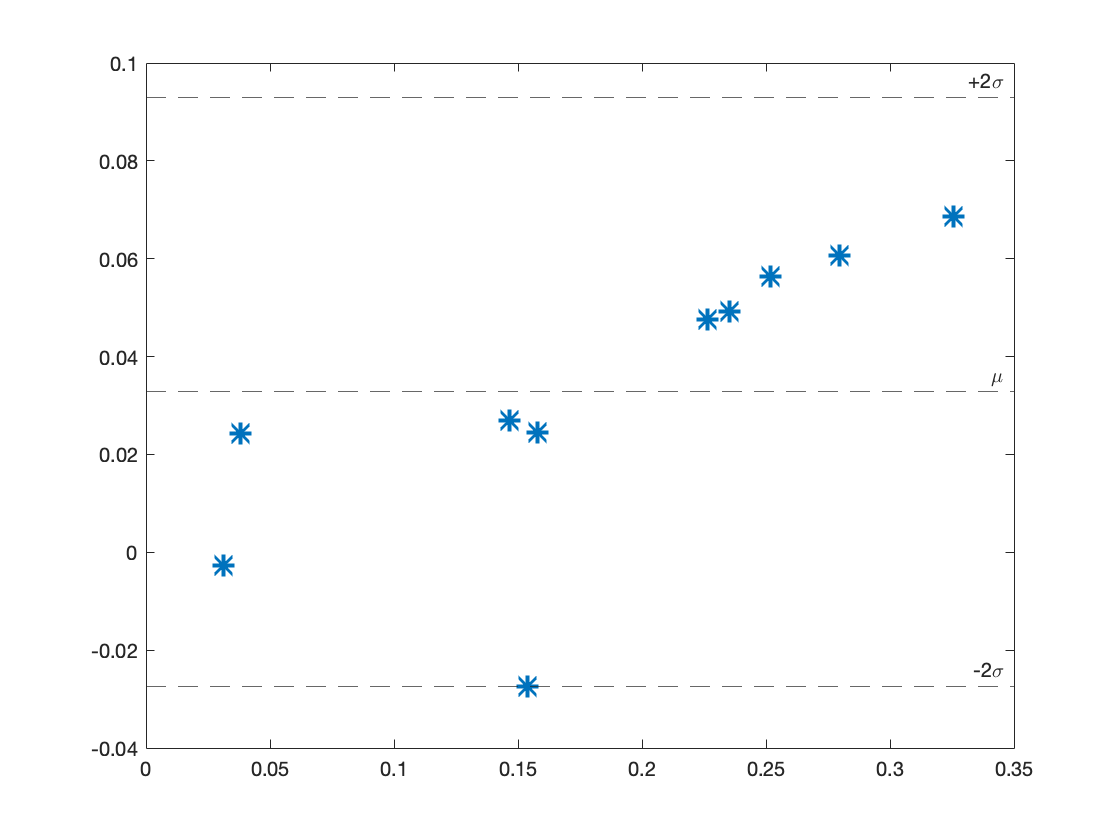}}
\\
\subfigure[FAZ segmentation (algorithm): \textit{perimeter }$(mm)$] {\includegraphics[width=0.45\textwidth]{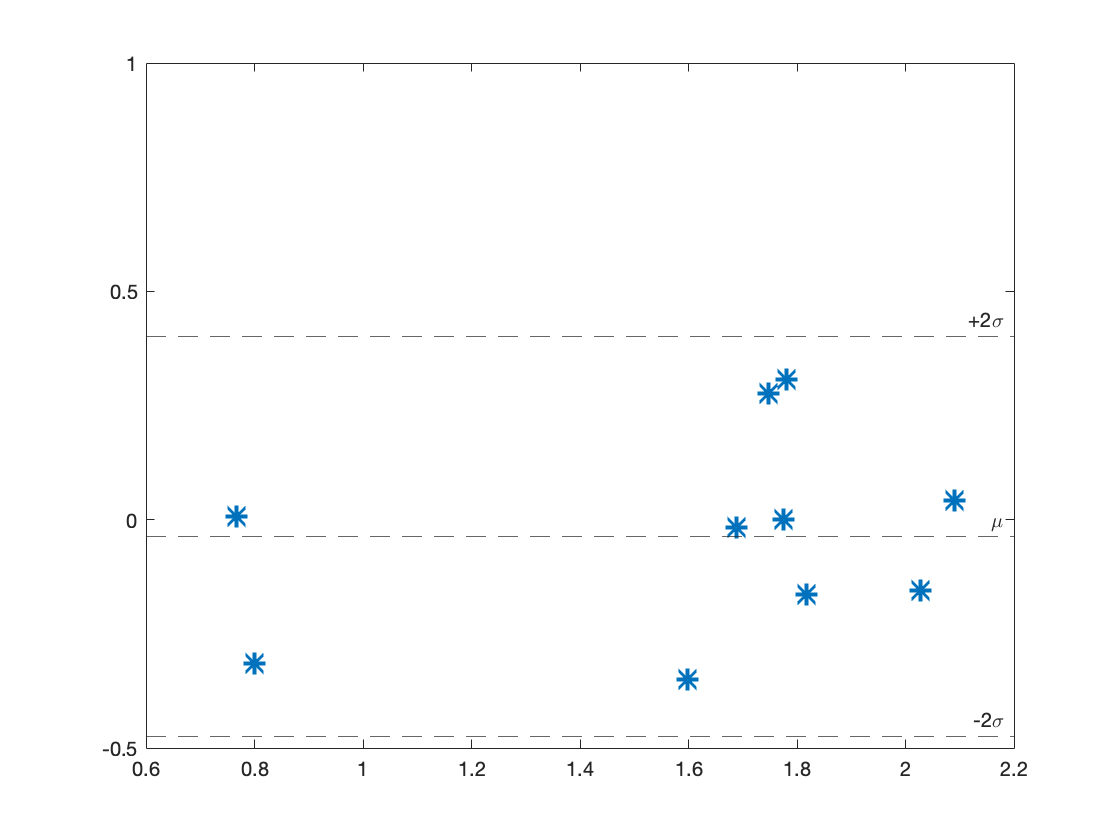}} 
\subfigure[FAZ segmentation (device): \textit{perimeter }$(mm)$] {\includegraphics[width=0.45\textwidth]{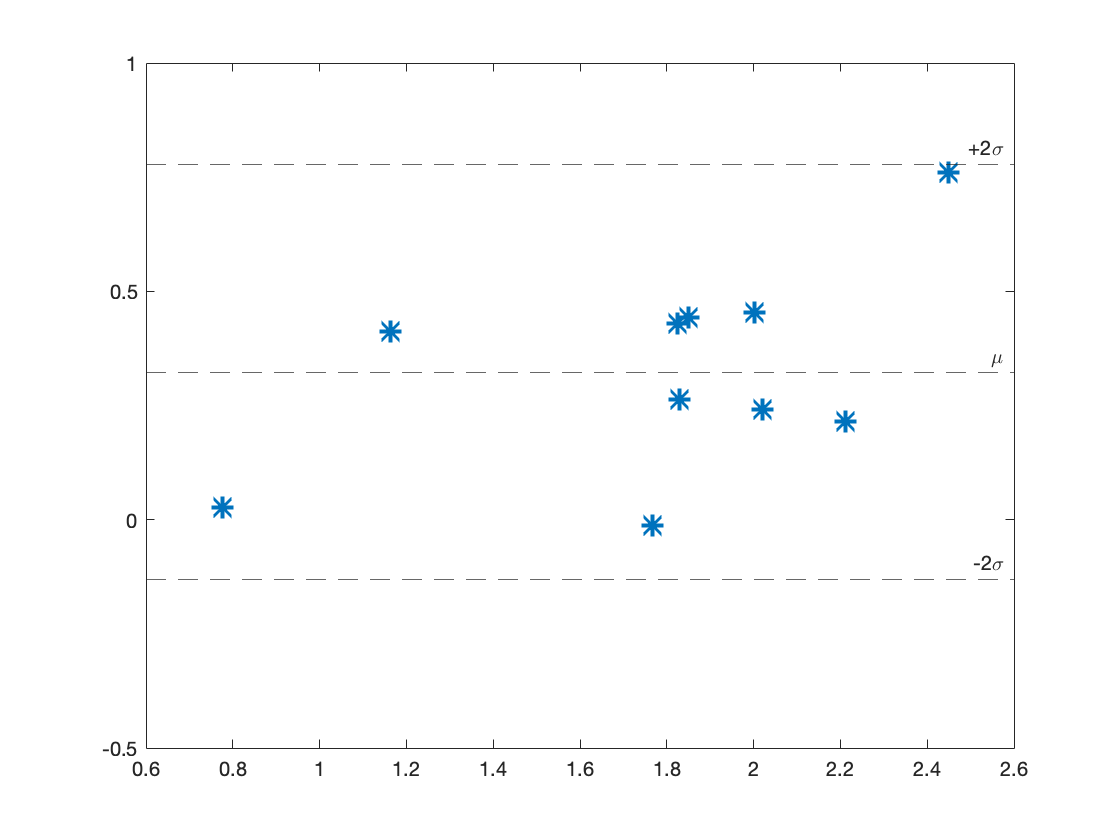}}
\\
\subfigure[FAZ segmentation (algorithm): \textit{circularity index}] {\includegraphics[width=0.45\textwidth]{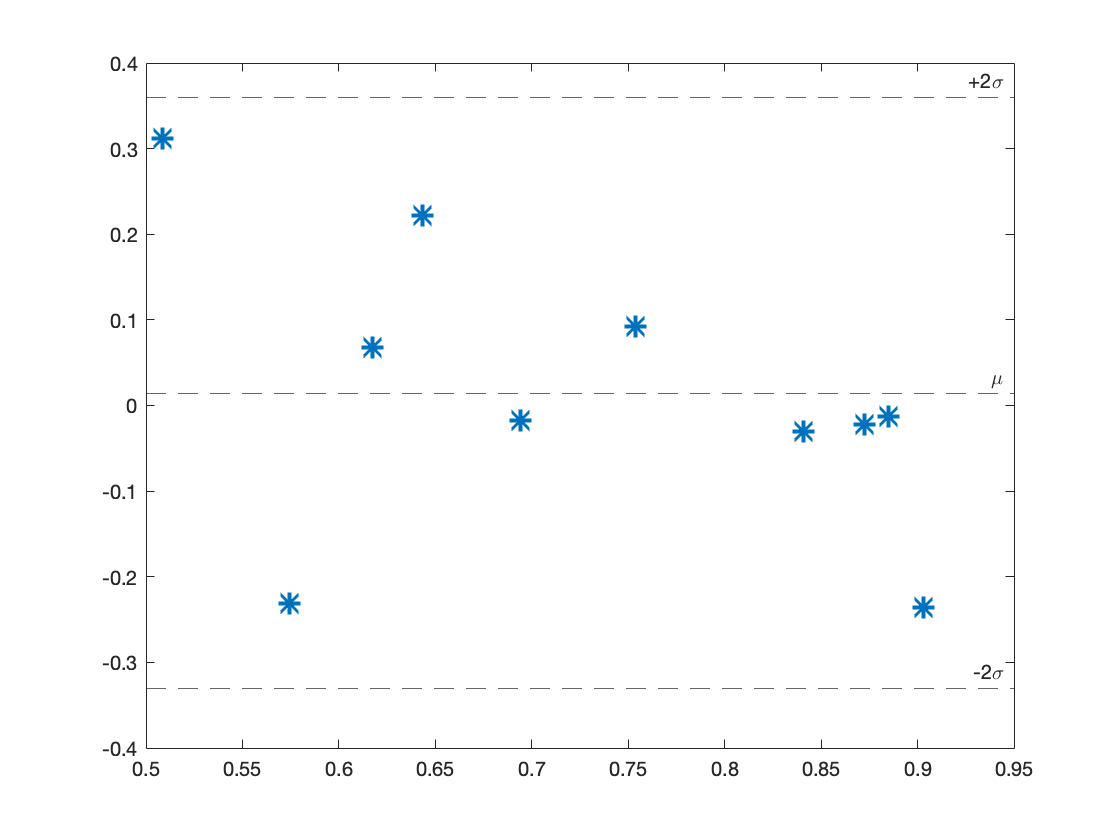}}
\subfigure[FAZ segmentation (device): \textit{circularity index}] {\includegraphics[width=0.45\textwidth]{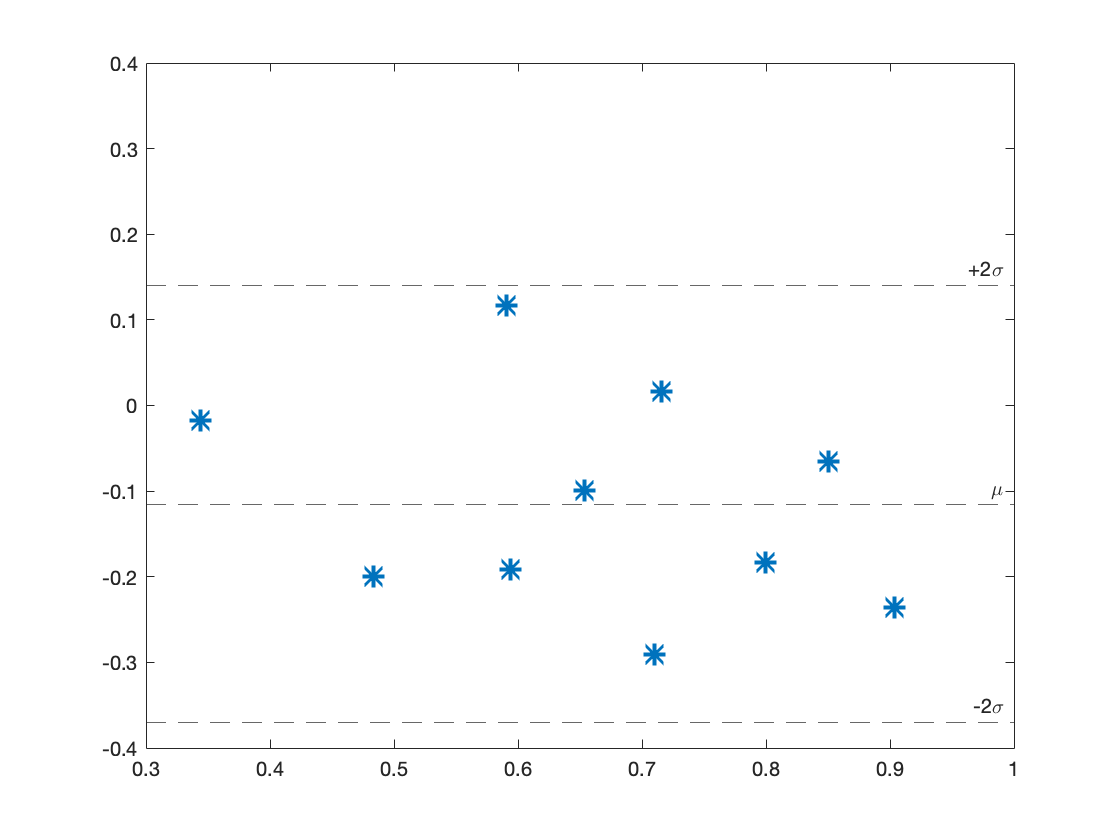}}

\caption{Bland-Altman Plots comparing the FAZ descriptors $area$, $perimeter$ and $circularity$ of the GT and our segmentations (left) and the automated output of the device (right). The $x$-axes denotes the mean of the observed values by the two methods, the $y$-axes the differences.}
\label{FazBA}
\end{figure}

The results when comparing the ground truth with our segmentations and the device's output are stated in Table \ref{FAZ}. We observe that the results of our algorithm coincide well with the ground truth and clearly outperform the FAZ segmentations provided by the device. We also computed the Bland-Altman plots for all $3$ quantities, which show a good agreement between our algorithm and the ground truth, see Figure \ref{FazBA}. In particular, it shows that all resulting differences are closely distributed around $0$ for our algorithm, whereas the output of the device differs much more from the GT regarding the quantity \textit{area}. The device's output is also biased towards yielding too big \textit{areas} and \textit{perimeters} as well as too little \textit{circularity}. 

\begin{table}
\centering
\caption{Absolute difference ($mean \pm SD$) of the FAZ identification compared to the GT.}
\label{FAZ}
\smallskip
\begin{tabular}{|c|c|c|c|}
\hline
& Area ($mm^2$) & Perimeter ($mm$) & Circularity \\
\hline
Device & $0.0388 \pm 0.0207$& $0.3256\pm 0.2233$ & $0.1416 \pm 0.0931$\\
Algorithm & $0.0068 \pm 0.0043 $& $0.1632 \pm 0.1409$ &  $0.1242 \pm 0.1136 $\\
\hline
\end{tabular}
\bigskip
\end{table}

Finally, we computed the mean dice similarity coefficient between our resulting segmentations and the manual annotations, yielding a good score of $0.89$ for the first and second grader respectively. 

\subsubsection{Local vessel density maps}
\begin{figure}
    \subfigure[Locality = 10] {\includegraphics[width=0.24\textwidth]{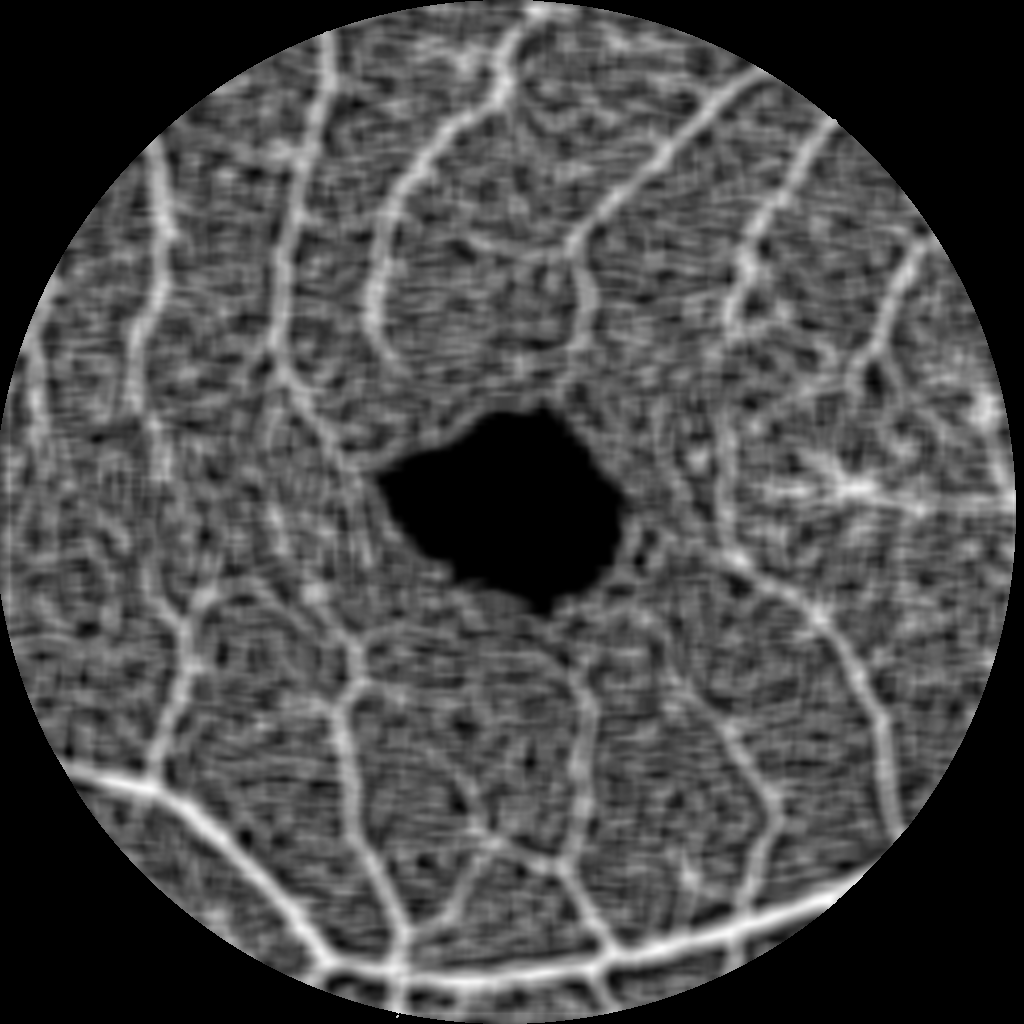}}
    \subfigure[Locality = 25] {\includegraphics[width=0.24\textwidth]{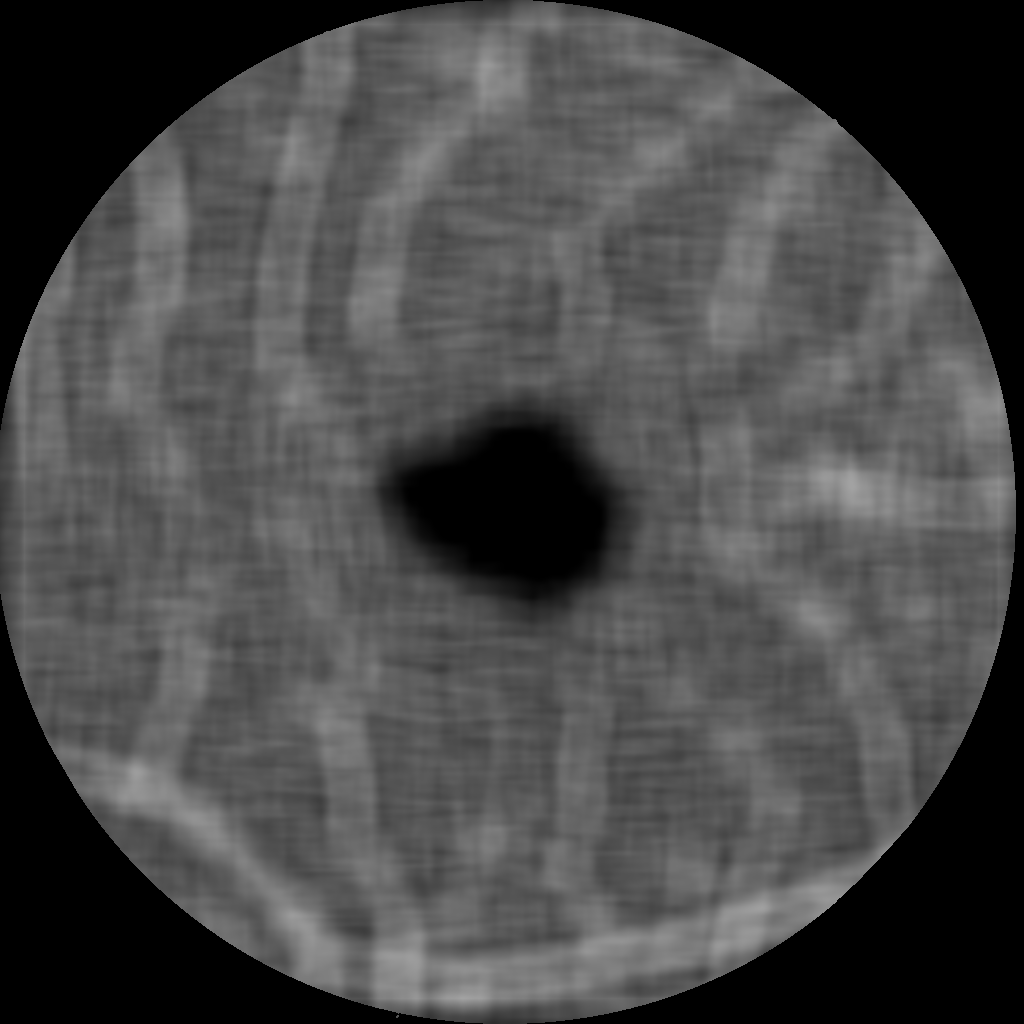} }
        \subfigure[Locality = 50] {\includegraphics[width=0.24\textwidth]{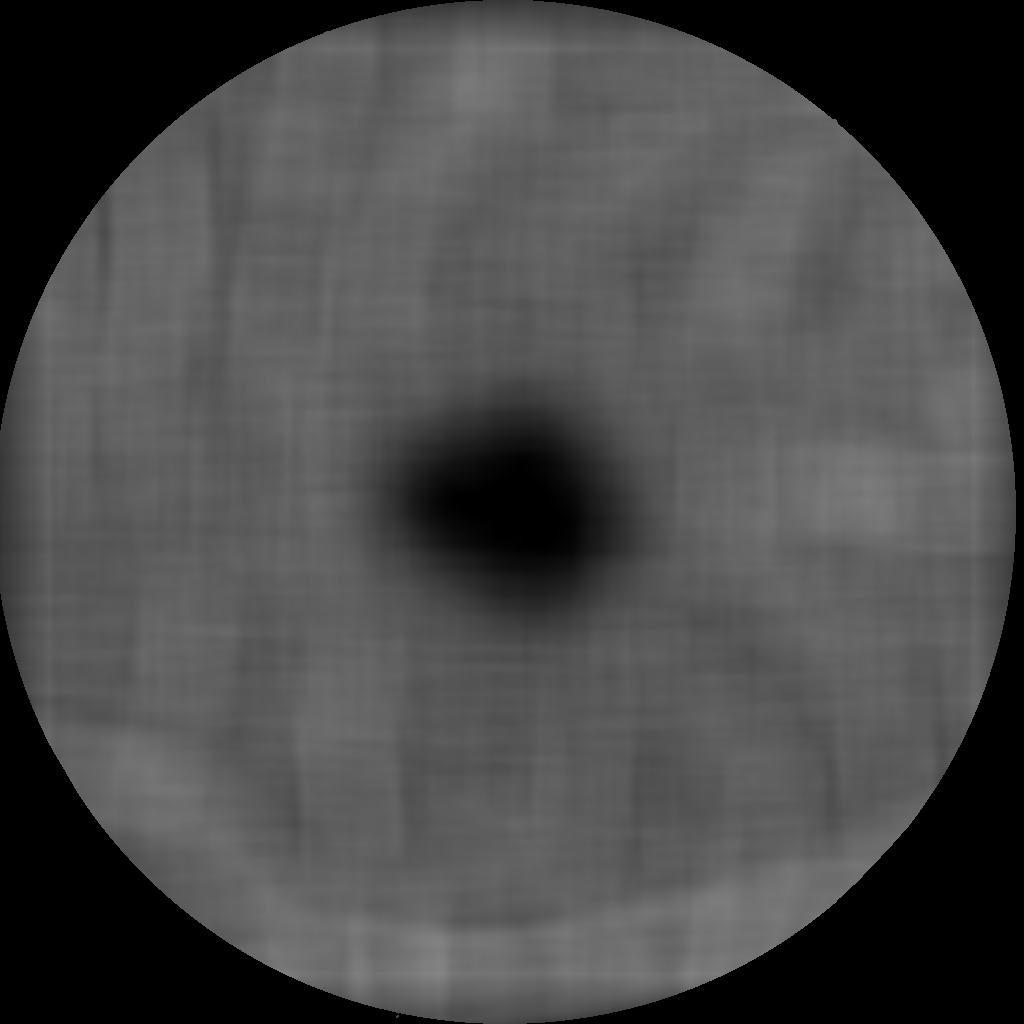}}
    \subfigure[Locality = 512] {\includegraphics[width=0.24\textwidth]{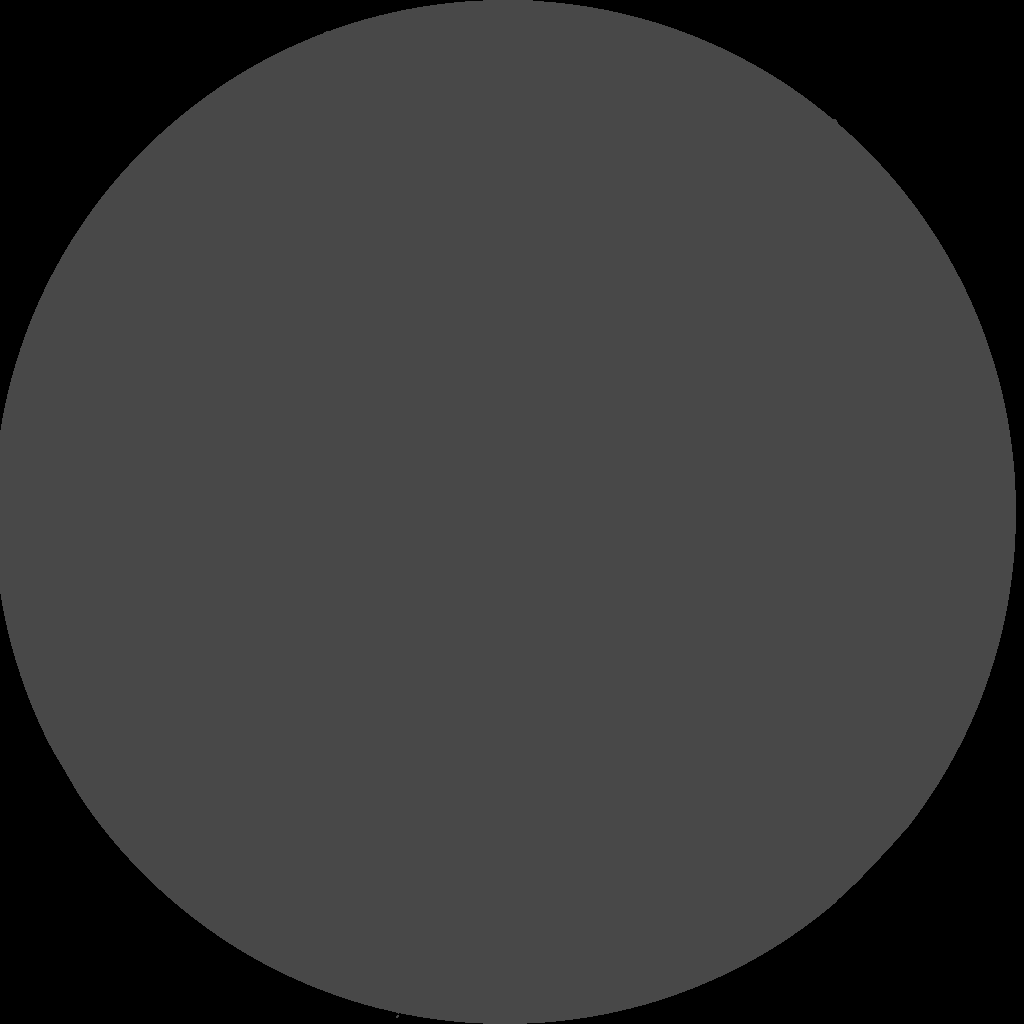} }
    \caption{Local VD maps for different locality sizes: the pixel values represent the VD in the area of its chosen pixel proximity. The locality chosen as the image size yields a constant map (d).}
    \label{localvd}
\end{figure}

Our method enables the computation of more localized VD values than what is provided by the device. Based on the accurate binary segmentation of the blood vessels, local VD maps could help clinicians to identify more inconspicuous regions of capillary dropout (e.g., in diabetes patients). We can provide VD maps with any predefined sizes of locality, see Figure \ref{localvd}, where brighter areas indicate high and darker areas less blood flow.

\subsubsection{Further perspectives} Our next planned steps are the evaluation of the method on data from different devices, including the deep capillary plexus, with diverse image quality as well as more detailed discussions with clinicians concerning pathologies. 
Moreover, we plan to adapt our algorithm to apply and compare it to recent publicly available data sets and algorithms\cite{10.1167/tvst.9.13.5,9284503}, including the computation of parameters from the vessel skeleton maps, see Figure \ref{VDbland}(b), that have been identified as other important parameters for quantitative assessment\cite{vdi}. 

\section{Conclusion}
We presented a novel algorithm for the segmentation of blood vessels of the superficial vascular network in OCTA scans, which is still a challenge for scientists in ophthalmology. The method is based on frequency filters that allow precise identification of smaller blood vessels. It yields convincing results visually as well as in quantitative comparison with in-built values from the OCTA device. The FAZ identification coincides very well with manual annotations by $2$ expert graders and even outperforms the automated segmentation provided by the device. Moreover, we observed that the VD in-built values are locally not consistent. Therefore, we suggest alternative representations of adaptive local VD maps based on our segmentation. Such maps might be helpful for clinical interpretation of various retinal and systemic diseases. Further experiments are planned including in-house data from different scanners and publicly available data sets involving manual annotations of the vessels. 
\bigskip 

\noindent
\textbf{Acknowledgements.} This work is funded by WWTF AugUniWien/FA7464A0249 \\ \mbox{(MedUniWien)} and VRG12-009 (UniWien). 
\bibliographystyle{abbrv}{}
\bibliography{filterocta.bib}

\end{document}